\documentclass{article}
\usepackage[]{acl}
\usepackage{url}
\usepackage{amsmath,amssymb,amsfonts}
\usepackage{xcolor}
\usepackage{booktabs}
\usepackage{multirow}
\usepackage{subcaption}
\usepackage{graphicx}
\usepackage{makecell}
\usepackage{dsfont}
\usepackage{threeparttable}
\usepackage{array}
\usepackage{tabularx}
\usepackage{diagbox}
\usepackage{arydshln}
\usepackage{float}
\usepackage{amsthm}
\usepackage{pgfplots}
\usepackage{tikz}
\usepackage{enumitem}
\usepackage{ragged2e}    

\definecolor{darkred}{RGB}{200, 0, 0}
\definecolor{darkyellow}{RGB}{250, 153, 0}
\definecolor{darkgreen}{RGB}{0, 180, 0}
\definecolor{lise}{RGB}{128,0,32}
\definecolor{deepblue}{RGB}{68,85,102}
\definecolor{graygreen}{RGB}{85,107,47}
\definecolor{tomato}{RGB}{255,99,71}
\definecolor{india}{RGB}{205,92,92}

\definecolor{saddlebrown}{RGB}{139, 69, 19}

\definecolor{forestgreen}{RGB}{34, 139, 34}
\definecolor{maroon}{RGB}{217,38,38}

\definecolor{navy}{RGB}{164, 101, 138}

\definecolor{moccasin}{RGB}{230, 159, 0}

\definecolor{darkred}{RGB}{200, 0, 0}
\definecolor{darkyellow}{RGB}{250, 153, 0}
\definecolor{darkgreen}{RGB}{0, 180, 0}

\title{Who Pays for Open Review? Visible Author Reputation and Its Effect on Ratings}

\author{
  \textbf{Qinghua Zhao}\textsuperscript{1},
  \textbf{Xinyu Chen}\textsuperscript{1},
  \textbf{Yanhui Yang}\textsuperscript{1},
  \textbf{Tengfeng Sun}\textsuperscript{1}, \\
  \textbf{Junfeng Liu}\textsuperscript{2} 
  \textbf{Zhongfeng Kang}\textsuperscript{3} \\
  \textsuperscript{1}Hefei University \quad
  \textsuperscript{2}Pengcheng Laboratory 
  \textsuperscript{3}Lanzhou University \\
  \texttt{zhaoqh@hfuu.edu.cn} \quad
  \texttt{liujf@pcl.ac.cn} \quad
  \texttt{kangzf@lzu.edu.cn} \\
  \texttt{\{chenxinyu, yangyanhui, suntengfeng\}@stu.hfuu.edu.cn}
}

\begin{document}
\maketitle

\begin{abstract}
An OpenReview bug in November 2025 broke anonymity at several conferences and prompted calls for open review, which motivate us to ask what shifting from blind to open would mean for authors. Analyzing over 18,000 reviewed submissions to ICLR 2026, split into \textit{de facto} open and blind groups by arXiv preprint timing, we find that ratings rise with author reputation under both mechanisms, with a steeper slope under open review that is statistically significant, and that the open–blind difference is concentrated at the borderline ratings. The pattern holds across five reputation proxies (including institution, h-index, and citation count), three author-aggregation rules, and five definitions of the open window. A controlled simulation with five AI models as reviewers, holding the manuscript fixed and varying the author reputation, reproduces the effect. With claude-opus-5 as the reviewer, for example, rating rises by 0.5 points as the author moves from low to high reputation. 

\end{abstract}

\section{Introduction}
Most major artificial intelligence conferences run peer review double-blind, with author and reviewer identities hidden from each other. In November 2025 an OpenReview API bug broke that anonymity across ICLR, ACL, KDD, ICML, and other venues, exposing the linkages between papers, authors, and their assigned reviewers. The incident drew a wave of discussion on social media, and part of it called for making review fully open. This incident motivates us to measure whether and how far author visibility moves review ratings.

Prior work shows that reviewers favor high-reputation authors when identity is visible \citep{sun2022does, stelmakh2023cite, ross2023open, pataranutaporn2025llm, ye2025justice}, but existing evidence comes from randomized trials on a few hundred papers or from venues that do not publish their review records. How large the effect is, where on the rating scale it acts, and whether it survives when the manuscript is held fixed and only reputation varies remain open. We address the first two questions on the review record of ICLR 2026 and the third through a parallel simulation in which AI reviewers rate the same submissions under different stated reputations, allowing the observational and experimental findings to corroborate each other.

ICLR is the only top conference that opens the complete review
record of every submission, where others such as NeurIPS release
only a selected part. A large share of its authors post their
work to arXiv before review starts, which gives a natural split
between \textit{de facto} open and blind submissions, and their
OpenReview profiles linked to OpenAlex let us measure each
author's reputation from institution and publication record.

We begin by crawling every submission of ICLR 2026 from OpenReview, with its content, author list, reviews, and decision, and by linking each author to OpenAlex, from which we assign reputation proxies based on institution and publication record. Since the same review process must be observed with and without author identity, we split the submissions into \textit{de facto} open and blind groups by whether a preprint appeared on arXiv before the review started. Experimental results show a \textcolor{deepblue}{reputation effect}: ratings rise with author reputation under both blind and open review but more steeply under the latter, and that the difference concentrates at the borderline ratings. Under the institution proxy, the slope difference is significant ($p = 0.008$) for mean-author and max-author ($p=0.041$). The pattern holds across five reputation proxies (i.e., institution reputation, h-index, citation count, publication count, and i10-index), three author-aggregation rules (i.e., the reputation of the first author, the max author, and the mean author), five definitions of the open window (i.e., within one month, within three months, more than three months, and more than six months before review opens, as well as all preprints before review), and three input scopes (i.e., abstract alone, abstract with introduction, and full main body). To control for paper quality, we fix the paper content and vary only the author reputation. Specifically, we sample 600 submissions and test with claude-opus-4-8, claude-opus-5, claude-sonnet-5, gpt-5.5, and gpt-5.6-sol as reviewers, each rating the same paper once without an author line and once each with a low-, middle-, or high-reputation author.  The rise from low to high reputation ranges from 0.2 to 0.6 points across four of the five models, consistent with the \textcolor{deepblue}{reputation effect} observed in the review record.

This paper makes three contributions. First, we construct a dataset that links every submission to ICLR 2026 with author reputation proxies drawn from OpenReview profiles and OpenAlex records. Second, we use this dataset to measure the \textcolor{deepblue}{reputation effect} on the review record of a top conference and locate it at the borderline ratings where acceptance is decided. Third, we run a controlled simulation in which AI reviewers rate the same manuscripts under different stated reputations, and the experimental estimates are consistent with the observational findings. 

\section{Related Work}

\paragraph{Reputation bias in peer review}
A long line of work studies how author reputation shapes review outcomes. Settings that reveal author information favor high-reputation researchers and prestigious institutions \citep{blank1991effects, tomkins2017reviewer, okike2016single, tran2020open, manzoor2021uncovering}, while hiding author information reduces this advantage \citep{kern2022impact, sun2022does}. A paper's rating is sensitive to whatever the reviewer can see about it beyond its content, so a visible resubmission tag or a citation to the reviewer's own work moves the rating \citep{stelmakh2021prior, stelmakh2023cite}, and author reputation is one more such signal.  These studies establish that a reputation channel exists and estimate its size in specific settings. They either compare across venues or years, run randomized trials on a few hundred submissions within one venue, or measure the reputation gradient under a single mechanism. This paper contrasts the two mechanisms on over 18,000 submissions to one conference through a natural split, and locates where on the rating scale the effect acts.

\paragraph{Preprints as a natural split}
Some AI conferences allow authors to post their work to arXiv before or during review. Whether a preprint itself changes the review outcome must be settled first. Studying five years of ICLR, \citet{zhang2023early} estimate the causal effect of early arXiving on acceptance and find it small and similar across author reputation strata.  We therefore treat a preprint posted before the review starts as an approximate marker of whether author reputation is visible to the reviewer rather than as a randomized assignment, and split the submissions of one venue into open and blind groups accordingly.

\paragraph{AI as reviewers}
AI-written and AI-assisted reviews are now widespread at major venues \citep{latona2024ai}. Prior work studies whether AI reviewers reach decisions consistent with human ones and finds that they track paper quality but compress the rating range and favor polished text \citep{zhou2024llm}. When author reputation is visible, AI reviewers further tilt toward recognized authors and prestigious institutions \citep{pataranutaporn2025llm, ye2025justice}. AI experimental conditions are  controllable, with the manuscript unchanged and only author reputation varying, so these experiments can isolate the \textcolor{deepblue}{reputation effect} from paper quality. However, prior work does not compare its ratings with the human reviews of the same submissions. This paper runs the simulation on the same papers whose human reviews are observed, so that the two sets of estimates can be compared directly.

\section{Analytical Framework}
\label{sec:methodology}
We formalize the conjecture that revealing author identity changes how the
same paper is scored. A review instance consists of an author with
reputation $\pi_a \in [0,1]$, a paper of quality $q$, and a review mechanism
$M \in \{\text{blind}, \text{open}\}$. Reputable authors may simply write
better papers, so we let $\bar{q}(\pi_a) = \mathbb{E}[q \mid \pi_a]$ be
non-decreasing in $\pi_a$. Under blind review the reviewer does not observe
$\pi_a$ and prices the submission at the prior mean reputation $\bar{\pi}$
of the pool; under open review the reviewer observes $\pi_a$ and, if visible
standing earns deference, prices it at $\pi_a$:
\begin{equation}
\label{eq:bias}
\mathbb{E}[r_M \mid \pi_a] = \bar{q}(\pi_a) +
\textcolor{lise}{\gamma} \cdot \begin{cases} \bar{\pi} & M = \text{blind}\\
\pi_a & M = \text{open.}\end{cases}
\end{equation}
The premium that opening the mechanism grants an author is therefore
\begin{equation}
\label{eq:B}
\Delta(\pi_a) = \textcolor{lise}{\gamma} \cdot (\pi_a - \bar{\pi}),
\end{equation}
which is positive above $\bar{\pi}$ and negative below it, so a reputation
effect is a transfer from the less established author to the more
established one rather than a bonus to everyone.

Three consequences guide the analysis. First, $\bar{q}$ is common to both
mechanisms, so $\textcolor{lise}{\gamma}$ is identified by the difference of the two
reputation slopes, which we estimate in Section \ref{sec:exp-gradients} as the coefficient on the reputation × open interaction in a linear regression, and a rating gradient under blind review is expected and reflects content alone. Second, a shift of size $\textcolor{lise}{\gamma} \pi_a$ moves a
paper across a threshold in proportion to the density of ratings there, so
its effect on outcomes is largest where submissions are dense, near the
acceptance borderline. Third, $\bar{q}$ is held fixed only if the
manuscript is, which a controlled simulation enforces directly and an
observational split approximates.

\begin{table}[!t]
\centering
\resizebox{0.9\columnwidth}{!}{%
\begin{tabular}{lrr}
\toprule
 & \textbf{Count} & \textbf{Share} \\
\midrule
 \textit{w/} a final decision & 13,726 & 100.0\% \\
\quad accepted & 5,352 & 39.0\% \\
\quad rejected & 8,374 & 61.0\% \\
\addlinespace
Distinct authors  & 49,246 & 100.0\% \\
\quad \textit{w/}  ORCID & 21,842 & 44.4\% \\
\addlinespace
 \textit{w/} first author ORCID & 5,912 & 43.1\% \\
\quad accepted & 2,298 & 38.9\% \\
\quad rejected &  3,614 & 61.1\% \\
\addlinespace
 \textit{w/} $\geq$1 ORCID author & 12,094 & 88.1\% \\
\quad accepted & 4,776  & 39.5\% \\
\quad rejected & 7,318   & 60.5\% \\
\midrule
ORCID authors per paper& 3.15 & -- \\
\bottomrule
\end{tabular}%
}
\caption{ORCID coverage of the OpenReview--OpenAlex linkage over 13,726 decided submissions. }
\label{tab:orcid-coverage}
\end{table}

\section{Data and Measurement}

\subsection{The ICLR 2026 Corpus}
\label{sec:corpus}

We collect every submission to ICLR 2026 through the official OpenReview API (v2, venue \texttt{ICLR.cc/2026/Conference}). For each submission we retrieve the PDF, the author list, the official reviews, and the final decision, and we resolve every author to their OpenReview profile, from which we take the current institutional affiliation and the ORCID identifier when present.

The crawl returns 19,814 submissions. We first drop the 908 papers that were desk-rejected, since they never entered peer review, leaving 18,906 reviewed submissions. Of these,  5,063  were withdrawn before the decision and  carry official ratings but no accept/reject outcome, while the remaining 18,789 received ratings and 13,726 received a final decision. The two counts define our two analysis samples: analyses of ratings use all 18,789 reviewed submissions with ratings, and analyses of acceptance use the 13,726 ones with decisions. 

Reviewers score each submission on the six-level ICLR scale $\{0, 2, 4, 6, 8, 10\}$. We use the final rating of each official review, that is, the score standing after the author response and reviewer discussion, and define a paper's rating as the unweighted mean of its final official ratings.

 \begin{table}[!t] 
  \centering
  \begin{tabular}{lrr}
  \toprule
   & Count & Share \\
  \midrule
 \multicolumn{3}{l}{\textit{Title matching against arXiv}} \\
  \quad Exact match            & 7,797& 56.8\% \\
  \quad Fuzzy match            &   707 &  5.1\% \\
  \quad Total matched          & 8,504 & 61.9\% \\
  \quad No arXiv match         & 5,222 & 38.1\% \\
  \midrule
  \multicolumn{3}{l}{\textit{Timing of the first arXiv version}} \\
  \quad Before review starts        & 5,189 & 37.8\% \\
  \quad During review / decision    & 1,837 & 13.4\% \\
  \quad After decisions released    & 1,478 & 10.8\% \\
  \midrule
  \multicolumn{3}{l}{\textit{Mechanism classification}} \\
  \quad \textit{De facto} open      & 5,189 & 37.8\% \\
  \quad Ambiguous        & 1,837 & 13.4\% \\
  \quad \textit{De facto} blind     & 6,700 & 48.8\% \\
  \quad\quad no arXiv match         & 5,222 & 38.1\% \\
  \quad\quad posted after decisions & 1,478 & 10.8\% \\
  \bottomrule
  \end{tabular}
  \caption{ArXiv matching and mechanism classification over the 13,726 submissions with a final accept or reject decision. }
  \label{tab:open-blind-split}
  \end{table}

\subsection{Measuring Author Reputation} \label{sec:reputation}
Author reputation admits no single agreed metric and is instead quantified through proxies drawn from an author's publication record and institutional affiliation. We use five base measures. First is  institution reputation, which we derive from the affiliation recorded on the author's OpenReview profile. Every submitting author must maintain an OpenReview profile.
Four are bibliometric and come from OpenAlex: the h-index, citation count, publication count, and i10-index. The two platforms are joined through the ORCID identifier on that profile, which resolves the author to a unique OpenAlex record. All measures are taken at the time of data collection (May 2026).

\paragraph{Institution reputation} The first measure reads reputation off the author's current institution string on OpenReview and maps it onto a single ten-tier scale, on which tier 1 is the most prestigious and tier 10 the least prestigious or unranked. The mapping separates judgement from lookup. A language model (\texttt{claude-opus-5}) first classifies each distinct string as a university, a company, or other, and for a university returns only its official top-level English name (``Dept.\ of EECS, MIT'' resolves to ``Massachusetts Institute of Technology''). That name is then matched deterministically against the QS World University Rankings 2027. Ranks are binned by value, so that the top tier remains a small elite group: tiers 1--9 cover ranks $\le 10$, 11--25, 26--50, 51--100, 101--200, 201--350, 351--500, 501--700, and 701--1000, and tier 10 collects ranks above 1000 and unranked institutions. Companies cannot be placed on a university ranking,  we therefore ask the model for an industrial research-prestige tier on the same scale, judged by the strength of the organisation's AI/CS research output, and anchored with named examples at each tier (tier 1 for leading industrial AI labs such as Google DeepMind, OpenAI, and Meta FAIR; tier 2 for established corporate research labs such as Microsoft Research, NVIDIA Research, and IBM Research; tiers 4--5 for large technology firms with a smaller research presence; tiers 8--9 as the default for companies the model does not recognise). Hospitals, government agencies, national laboratories, and unidentifiable strings are classified as other and receive no value.  Of the 13,066 distinct institution strings, 5,957 are universities, 4,760 companies, and 2,349 other, and 10,717 (82.0\%) receive a tier; at the author level this covers 101,574 of 112,580 authorships (90.2\%), the broadest coverage of any proxy. In all analyses the tier enters negated, so that a higher value means a more reputable author and a positive slope means reputation is rewarded. The tier is ordinal and we treat it as cardinal only to report an interpretable slope.

 \paragraph{Bibliometric measures} Linkage through ORCID is the binding constraint on coverage. Table~\ref{tab:orcid-coverage} reports it over the decided submissions: 44.4\% of the 49,246 distinct authors carry an ORCID, and 88.1\% of submissions have at least one such author.  Of the 5,912 submissions whose first author records an ORCID, only  2,626 (44.4\%) resolve to an OpenAlex author record, the remaining 4,258 ORCIDs return no record, whereas every resolved author has a non-missing h-index. The bibliometric measures are therefore available for a minority of first authors but for at least one author on most papers.

  \begin{table}[!t]
\centering
\small
\begin{tabular}{@{}lp{0.72\linewidth}@{}}
\toprule
Level & Natural Language Description \\
\midrule
Low & a less established researcher with limited prior publications \\
Middle & a mid-career researcher with a moderate publication record \\
High & a senior researcher well known in the field, with over 100{,}000 citations \\
\bottomrule
\end{tabular}
\caption{{The three author reputation levels and the description attached to the submission under open review, quoted verbatim from the prompt.}}
\label{tab:rep_levels}
\end{table}

\subsection{Labelling Submissions as De Facto Open or Blind} \label{sec:split}
To study how the review mechanism impacts ratings under different author reputation, we need to divide submissions into those reviewed with author identity visible and those reviewed with identity hidden. ICLR provides no such division explicitly, but permits authors to post their work to arXiv, and a reviewer of a preposted paper could in practice recover its authors from the public version. We therefore classify each submission as \textit{de facto} open or \textit{de facto} blind (that is, open or blind in practice though not by official designation) by the timing of its first arXiv version.

\begin{figure*}[!t]
\centering
\includegraphics[width=\linewidth]{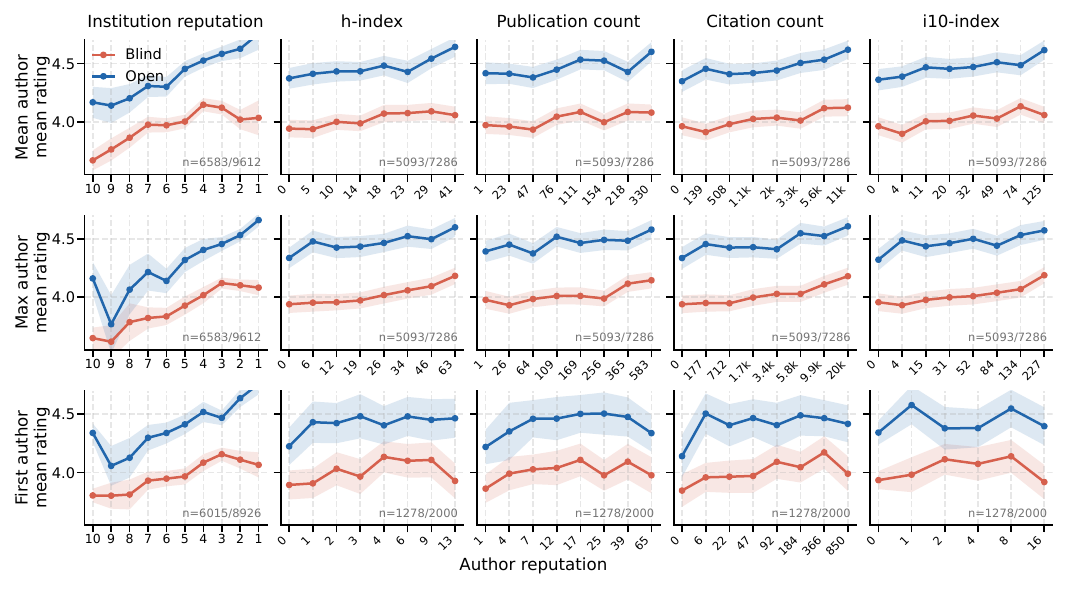}
\caption{Author reputation and mean review rating in the \textit{de facto} blind (red) and open (blue) reviewed papers, for five reputation proxies $\pi_a$ (columns) under three author-aggregation rules (rows), with 95\% intervals. Panel insets give the open and blind sample sizes. Other parameters are kept default.}
\label{fig:rating_by_reputation}
\end{figure*}
Specifically, starting from the submissions in which at least one author carries an ORCID (Table~\ref{tab:orcid-coverage}), we link each to arXiv against a local copy of the Cornell arXiv metadata snapshot (Kaggle \texttt{Cornell-University/arxiv}), downloaded 2026-05-14, containing 3,034,697 records with metadata updated through 2026-05-08, so that every submission is checked against the same frozen index. To match a submission to its arXiv counterpart, we normalize both titles and compute the proportion of overlapping words, and further require  that the first author's surname and first-name initial agree exactly, so that ``M\"{u}ller'' matches ``Muller''.

We then classify each matched submission by the timing of its first arXiv version against two instants of the ICLR 2026 timeline: the opening of the review period at 2025-10-10 00:00 UTC, and the release of decisions at 12:00 AoE on January 25, 2026. A submission whose first arXiv version predates the former is labelled \textit{de facto} open, whereas a submission with no arXiv version before the latter is labelled \textit{de facto} blind, and a submission whose first version falls between the two is ambiguous and therefore set aside. In addition, we further test four narrower definitions of open, under which a submission counts as open only if its first arXiv version appeared within 1 month of the review start, or within 3 months of it, or more than 3 months before it, or more than 6 months before it. Table~\ref{tab:open-blind-split} reports the resulting partition.

\section{Empirical Evidence}
\label{sec:evidence}

\begin{table}[!t]
\centering
\resizebox{\columnwidth}{!}{%
\begin{tabular}{lccc}
\toprule
Aggregation & Blind slope & Open slope & $\textcolor{maroon}{p}$ \\
\midrule
First author & 0.042 & 0.049 & 0.306 \\
Max author   & 0.057 & 0.072 & \textcolor{maroon}{0.041} \\
Mean author  & 0.050 & 0.071 & \textcolor{maroon}{0.008} \\
\bottomrule
\end{tabular}%
}
\caption{$p$ value of $t$-test  of the open--blind slope difference $\textcolor{lise}{\gamma}$ on  institution reputation proxy.}
\label{tab:gamma_test}
\end{table}

\subsection{\textcolor{forestgreen}{Quantifying the reputation effect}}
\label{sec:exp-gradients}
To measure how far author reputation moves the rating under each review
mechanism, we compare the rating of open and blind papers. Specifically, we
partition papers into equal-frequency reputation bins so that each bin
contains a roughly comparable number of submissions. Next, for each
reputation proxy, we compute the  slope of the rating on the log
reputation, and take the difference of the two slopes between open and blind
reviewed papers as the \textcolor{deepblue}{reputation effect} $\textcolor{lise}{\gamma}$ defined in the analytical
framework. To test whether $\gamma$ differs from zero, we fit the regression and report the $t$-test with $p$ value of the interaction coefficient.
Sample size varies across proxies. The bibliometric proxies require an ORCID
resolution to OpenAlex and cover a minority of first authors, whereas the
institution proxy comes directly from OpenReview profiles and covers the
largest sample, so we treat it as our primary analysis.

Figure~\ref{fig:rating_by_reputation} shows that under most reputation proxies, the mean rating rises with reputation, but the open line climbs more steeply than the blind one. This pattern is most pronounced under the institution proxy. The open and blind slopes of the mean rating on negated institution tier are $0.049$ and $0.042$ (first-author), $0.072$ and $0.057$ (max-author), and $0.071$ and $0.050$ (mean-author), giving $\textcolor{lise}{\gamma}=0.007$, $0.015$, and $0.021$, which amounts to about $0.06$ to $0.19$ rating points over the observed reputation range. Table~\ref{tab:gamma_test} reports the interaction test: $\gamma$ is significant under mean-author aggregation ($p = 0.008$) and max-author ($p = 0.041$). The other proxies show the same direction with a weaker open--blind separation. We do not test the slope difference for these proxies, since their ORCID-dependent coverage is far smaller and the estimates would rest on a selected subsample.
 ORCID linkage to OpenAlex is more reliably populated for established researchers, so these samples are already filtered toward higher reputation and the contrast is estimated over a compressed range. We next ask whether the \textcolor{deepblue}{reputation effect} is spread evenly over the rating scale or concentrated in a particular range.

\begin{figure*}[!t]
\centering
\includegraphics[width=\linewidth]{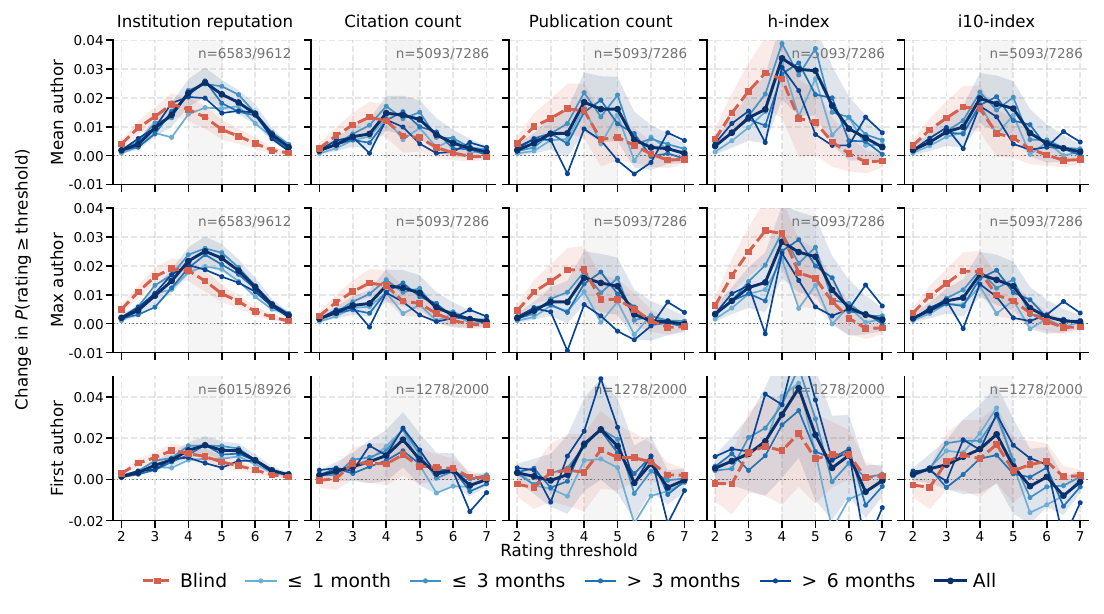}
\caption{Effect of author reputation on whether a paper reaches each level of the rating scale, shown separately for the de facto blind (red) and open (blue) subsamples across five reputation measures (columns), three author-aggregation rules (rows), and five arXiv timing windows (blue curves). Shaded bands give 95\% intervals for the blind and the pooled open series, panels in each row share the same vertical scale, and panel insets give the open and blind sample sizes.}
\label{fig:threshold-slopes}
\end{figure*}

\subsection{\textcolor{darkyellow}{Where the Reputation Effect Lands}}

\label{sec:exp-thresholds}

We next ask where on the rating scale the \textcolor{deepblue}{reputation effect} lands, and in particular whether it is concentrated at the borderline that decides acceptance. For each threshold $c$ on the rating scale, we regress the indicator $\mathbf{1}\{\text{rating} \ge c\}$ on author reputation and report the estimated coefficient, separately for the open and the blind reviewed papers. Plotting these coefficients against $c$ traces out where a unit of reputation shifts a paper across the bar and where it does not.

Figure~\ref{fig:threshold-slopes} shows that across all evaluated scenarios, the overall trend consistently exhibits an inverted U-shape. As the rating threshold increases, the change in the probability of reaching it initially rises,  peaks between ratings $3.5$ and $4.5$, and then declines continuously. Taking the institution proxy (max author) as an example, among open reviewed papers the change rises from $0.002$ at a threshold of $2.0$ to $0.025$ at $4.5$, and falls back to $0.003$ at $7.0$. The peak thus sits in the borderline region where acceptance is decided, which is where the \textcolor{deepblue}{reputation effect} does most of its work.

The two curves differ in where they peak. The blind curve peaks at a threshold of about $3.5$ and the open curve at about $4.5$, so under open review the \textcolor{deepblue}{reputation effect} is largest closer to the level at which acceptance is decided. At the extremes of the scale the two curves nearly coincide, so reputation neither rescues a clear reject nor is needed for a clear accept. But author reputation is entangled with paper quality, and the observational split cannot tell the two apart. We therefore turn to a setting where the manuscript is held fixed and only the stated identity varies.

\begin{figure}[!t]
    \centering
    \includegraphics[width=\linewidth]{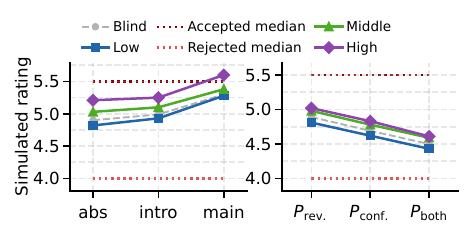}
    \caption{Mean simulated rating averaged over the five models, varying the
input under $P_{\text{both}}$ (left) and the prompt under the default input
(right), by author reputation condition. }
\label{fig:ablation}
\end{figure}

\subsection{\textcolor{india}{Isolating Identity from Content}}
\label{sec:llm_experiment}
The gap measured above may reflect reputation or the quality. We therefore place AI models in the reviewer role, where the
same submission can be rated repeatedly under different author reputation
with the manuscript held fixed.
\begin{figure*}[!t]\centering
\includegraphics[width=\linewidth]{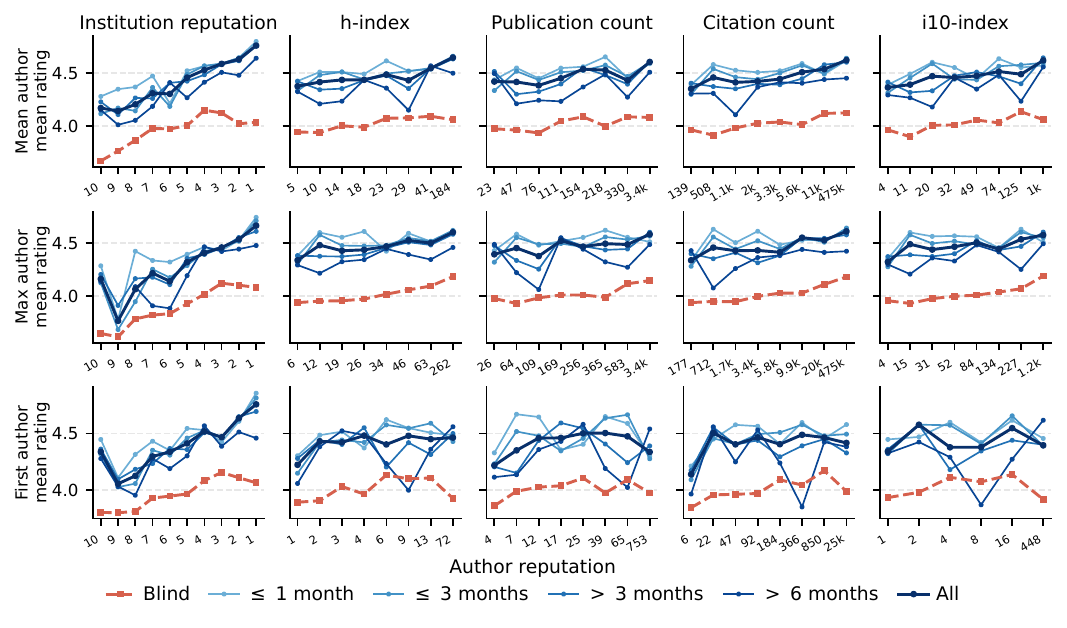}
\caption{Mean review rating by reputation bin for the
\textit{de facto} blind subsample (dashed) and for the open
subsample under five definitions of openness.}
\label{fig:rating_windows}
\end{figure*}

\paragraph{Setup}
We sample 600 reviewed submissions, stratified by their real mean rating in
bins of 0.5 and drawn in proportion to each bin's share of the corpus, with
a floor of 60 submissions per bin so that the whole scale is covered. Each
review takes as input the content of the submission and a description of its
author's reputation, rendered as one of the three natural-language levels in
Table~\ref{tab:rep_levels}. Every submission is reviewed under four conditions,
namely blind and open with a low-, middle-, or high-reputation author, by
five frontier models from two developers: claude-opus-4-8, claude-opus-5,
claude-sonnet-5, gpt-5.5, and gpt-5.6-sol. The AI reviewer returns a rating on
the ICLR scale of $\{0,2,4,6,8,10\}$. Under open review the reputation description accompanies the submission and
under blind review it is omitted, with everything else unchanged. We use three review instructions.
$P_{\text{conference}}$ (Table~\ref{tab:llm_prompt_p1}) describes the conference and marks which points of
the scale count as rejection and which as acceptance.
$P_{\text{reviewer}}$ (Table~\ref{tab:llm_prompt_p2}) casts the reviewer as a senior expert and lists six
aspects of judgement, from presentation and experiments to clarity,
soundness, and the core idea. $P_{\text{both}}$ (Table~\ref{fig:prompt-p3-text}, \ref{fig:prompt-p3-image}) combines the two and anchors
every point of the scale. Unless otherwise stated, the default configuration
supplies the abstract and the introduction under $P_{\text{both}}$.

\begin{figure*}[!t]
\centering
\includegraphics[width=\linewidth]{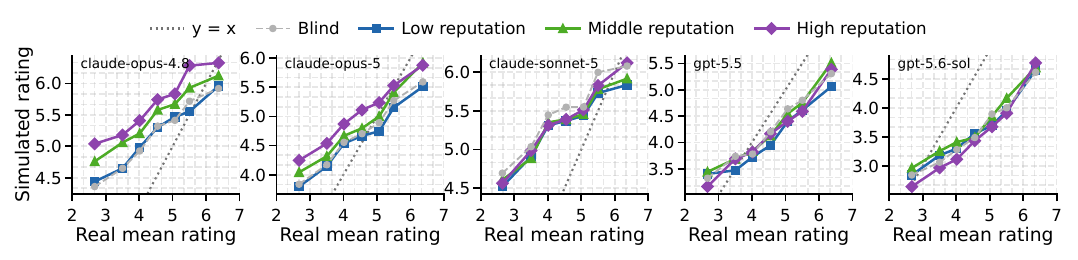}
\caption{Mean simulated rating against the real mean rating under the default setting ($P_\text{both}$ prompt, abstract and introduction as input), by author reputation condition. The dotted diagonal marks $y=x$: points above it are rated higher than by the human reviewers.}
\label{fig:llm-sim-results}
\end{figure*}

\paragraph{The \textcolor{deepblue}{reputation effect}}
Figure~\ref{fig:llm-sim-results} shows the result. In four of the five
models, all except gpt-5.6-sol, the same paper receives a higher rating as
the stated author moves from the low- to the high-reputation condition. The
rise lies between 0.2 and 0.6 points across the four models, and reaches 0.5
points in claude-opus-5. Since the manuscript is held fixed, the rise cannot come from differences in
content, and the stated reputation is the most likely source, which is
consistent with the observational estimates.

\begin{figure*}[!t]\centering
\includegraphics[width=0.9\linewidth]{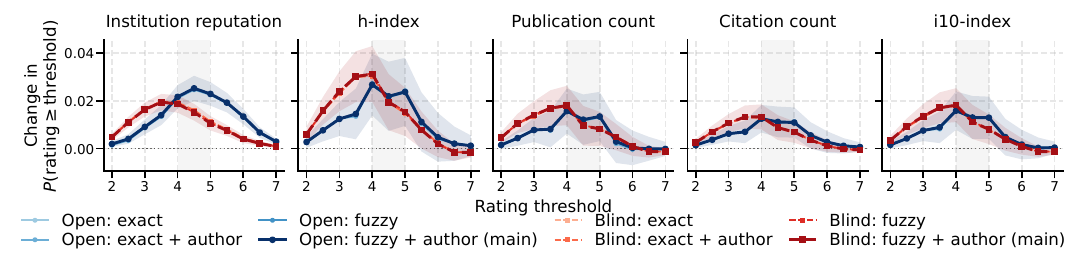}
\caption{Effect of author reputation on the probability of
reaching each rating threshold under the four ICLR--arXiv
matching rules, with the darkest shade marking the main
fuzzy-with-author-check rule.}
\label{fig:match-sweep}
\end{figure*}

\paragraph{Varying the  input}
The default input may be too thin for the reviewer to separate submissions
on content, so we rerun the simulation under $P_{\text{both}}$ with three
inputs, namely the abstract alone, the abstract with the introduction, and
the full main body supplied as rendered page images. The left panel of
Figure~\ref{fig:ablation} reports the mean rating averaged over the five
models under these three inputs. Across all of them the four conditions keep
the same ordering, with the high-reputation author above  the low-reputation author below it, and the high$-$low gap stays
positive ($0.39$, $0.32$, and $0.32$). The \textcolor{deepblue}{reputation effect} is therefore insensitive to the amount of the paper
shown to the reviewer.

\paragraph{Varying the prompt}
The wording of the prompt may itself steer the rating, so we repeat the
simulation with the input fixed to the abstract and the introduction under
the three prompts $P_{\text{conference}}$, $P_{\text{reviewer}}$, and
$P_{\text{both}}$. The right panel of Figure~\ref{fig:ablation} reports the
mean rating averaged over the five models under these three prompts. Across
all of them the four conditions keep the same ordering, with the
high-reputation condition on top and the low-reputation condition at the
bottom, and the high$-$low gap stays positive ($0.21$, $0.21$, and
$0.18$). The \textcolor{deepblue}{reputation effect} is therefore insensitive to the wording of
the prompt.

\paragraph{Comparing simulated and real ratings}
Two further observations follow from Figure~\ref{fig:llm-sim-results}.
First, the simulated ratings track the real ones, with a correlation of
about 0.35 in every model and under both mechanisms, so the AI reviewers
recover paper quality to some degree from the abstract and the introduction
alone. Second, the simulated ratings are compressed toward the middle of the
scale. Submissions rated 2.7 by the human reviewers receive between 4.3 and
5.6, and those rated 6.4 receive between 5.7 and 6.5.

\subsection{\textcolor{tomato}{Robustness to Data Construction}}
\label{sec:robust-data}
The \textcolor{deepblue}{reputation effect} can be influenced by the labelling of a submission as
open, which follows from the arXiv window that counts as open and from the
rule that matches a submission to its arXiv record. We vary both and test
the robustness of the effect.

\paragraph{Varying the open window}
The main analysis counts a submission as open whenever its first arXiv version predates the review start. We recode openness under four narrower definitions, requiring the first version to fall within one month or within three months of the review start, or to predate it by more than three or more than six months. Figure~\ref{fig:rating_windows} shows that under all five definitions the open line climbs more steeply than the blind line, as in Figure~\ref{fig:rating_by_reputation}. The \textcolor{deepblue}{reputation effect} is thus consistent across the open windows.

\paragraph{Varying the matching rule}
The main analysis links a submission to arXiv by fuzzy title matching with a first-author check. We rerun the pipeline under the other three rules, which match titles exactly instead of fuzzily, drop the author check, or do both. Figure~\ref{fig:match-sweep} shows the results. Under all four rules the effect peaks in the same borderline region and the open profile stays above the blind one, as in Figure~\ref{fig:threshold-slopes}. The \textcolor{deepblue}{reputation effect} is thus consistent across the matching rules.

\begin{figure}[!t]
\centering
\includegraphics[width=0.8\linewidth]{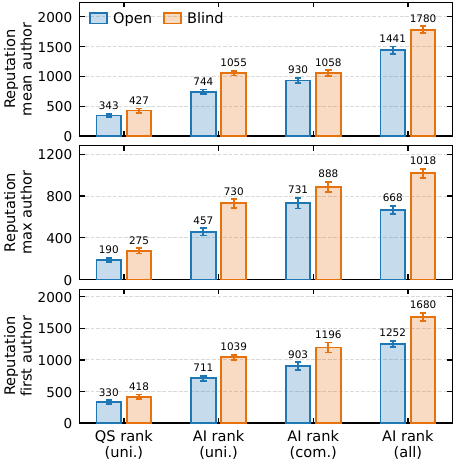}
\caption{Across all twelve combinations of four reputation measures
and three author-aggregation rules, authors of \textit{de facto} open
papers are  more reputable than authors of blind papers.}
\label{fig:reputation-selection}
\end{figure}

\section{Conclusion}
This paper studies how author reputation affects the review rating once the reviewer can see who wrote the paper. On the complete review record of ICLR 2026, split into \textit{de facto} open and blind submissions by arXiv preprint timing, the rating rises with author reputation under both mechanisms but more steeply under open review, significantly so under mean-author aggregation, and the difference concentrates at the borderline ratings that decide acceptance. A controlled simulation with AI reviewers, in which the manuscript is fixed and only the author reputation varies, shows the same trend and separates reputation from content. Anonymity thus reduces the extent to which visible standing, rather than content, moves a rating.

\section*{Limitations}

\paragraph{\textcolor{saddlebrown}{Bias between open and blind}}
We use arXiv preposting as a proxy for author
identity visibility. Authors who prepost are not randomly drawn from
the submission pool. To assess whether this confound is material, we
compare the institutional reputation of the open and blind groups
across four independent reputation measures and three
author-aggregation rules, as shown in Figure~\ref{fig:reputation-selection}. The four measures are the
QS World University Ranking restricted to university-affiliated
authors, an independent LLM-assigned reputation rank restricted to
university-affiliated authors, the same LLM-assigned rank restricted
to company-affiliated authors, and the two LLM rank pools combined so
that no institution type is excluded. 
Under the QS
measure alone, mean ranks are 343 against 427 for the mean-author
aggregation. The direction is
identical under all three LLM-based measures, and the gap is
comparable in magnitude. Across all twelve combinations,
\textcolor{saddlebrown}{open-group authors consistently hold a lower rank number, indicating
higher institutional reputation, than blind-group authors}, and
the observational rating gap should be read as an upper bound on the
\textcolor{deepblue}{reputation effect} rather than a  causal estimate. We accordingly
treat the observational analysis as descriptive evidence of scale and
direction.

\paragraph{\textcolor{saddlebrown}{Reputation proxies}}
Our bibliometric proxies are measured at the time of data collection
in May 2026 rather than at the time of review, because h-index,
citation count, and publication count are retrieved from OpenAlex as
present-day snapshots. Since these metrics are monotonically
non-decreasing over time, the lag understates rather than overstates
authors' reputation at the time of review, and any resulting bias is
conservative and cannot reverse the sign of the reported effects.

\paragraph{\textcolor{saddlebrown}{AI Reviewer model}}
Using models with an older knowledge cutoff would eliminate the risk
of the model having seen these submissions during pretraining, but \textcolor{saddlebrown}{
older models usually rate every paper a high rating and would be poor stand-ins for real reviewers.} We
therefore use current models and address contamination through study
design: since each paper is scored under multiple author identity
conditions, any pretraining exposure affects all conditions equally
and does not confound the within-paper contrast we report.

\paragraph{\textcolor{saddlebrown}{Evidence chain.}}
None of the limitations above undermines the central conclusion,
because the \textcolor{saddlebrown}{reputation effect does not rest on any single piece of
evidence for its causal support but on a chain of evidence that
converges from several independent directions}.  The 18,000 submissions establish the direction and rough size of the effect, with the slope difference reaching significance under mean-author aggregation and marginal significance under max-author. The controlled simulation reproduces the effect with the manuscript held fixed, closing the gap. The pattern further holds across 5 reputation proxies, 3
author-aggregation rules, 5 open-window definitions, and 4
arXiv-matching rules. 

\bibliography{tacl2021}

\appendix

\clearpage

\section*{Appendix} \label{app:prompts}

\begin{table}[!ht]
\centering
\small
\begin{tabular}{@{}p{0.96\linewidth}@{}}
\toprule
\texttt{You are reviewing a submission to ICLR 2026. ICLR is a top-tier machine learning conference. A paper accepted at ICLR is expected to present a novel idea that advances the field, to support its claims with thorough experiments that include strong baselines and ablations, and to be written clearly enough that a reader can follow the contribution and reproduce the work.} \\
\addlinespace
\texttt{Title: \{paper\_title\}} \\
\texttt{Author: \{author\_reputation\_descriptor\}.} \\
\addlinespace
\texttt{Abstract: \{paper\_abstract\}} \\
\texttt{Introduction: \{paper\_introduction\}} \\
\addlinespace
\texttt{Assign an overall rating on the ICLR 2026 scale. Ratings of 0, 2, and 4 are negative and mean the paper should be rejected; ratings of 6, 8, and 10 are positive and mean the paper should be accepted:} \\
\texttt{10 = strong accept} \\
\texttt{8 = accept} \\
\texttt{6 = weak accept} \\
\texttt{4 = borderline reject} \\
\texttt{2 = reject} \\
\texttt{0 = strong reject} \\
\addlinespace
\texttt{Respond with JSON only:} \\
\texttt{\{"rating": <one of 0, 2, 4, 6, 8, 10>, "confidence": <integer 1-5>, "justification": "<at most 50 words>"\}} \\
\bottomrule
\end{tabular}
\caption{The \textcolor{darkyellow}{$P_{\text{conference}}$} prompt, quoted verbatim, shown under open review with the \textcolor{darkyellow}{abstract and the introduction} as the content. The Author line is omitted under blind review and the prompt is otherwise byte-identical.}
\label{tab:llm_prompt_p1}
\end{table}

\begin{table}[!ht]
\centering
\small
\begin{tabular}{@{}p{0.96\linewidth}@{}}
\toprule
\texttt{You are reviewing a submission to ICLR 2026. You are a senior expert with a background in artificial intelligence, machine learning, natural language processing, and computer vision. Judge the submission on its presentation, the extent of its experimental evaluation, the quality of its writing, its clarity, its soundness, and the strength of its core idea.} \\
\addlinespace
\texttt{Title: \{paper\_title\}} \\
\texttt{Author: \{author\_reputation\_descriptor\}.} \\
\addlinespace
\texttt{Abstract: \{paper\_abstract\}} \\
\texttt{Introduction: \{paper\_introduction\}} \\
\addlinespace
\texttt{Assign an overall rating on the ICLR 2026 scale:} \\
\texttt{10 = strong accept} \\
\texttt{8 = accept} \\
\texttt{6 = weak accept} \\
\texttt{4 = borderline reject} \\
\texttt{2 = reject} \\
\texttt{0 = strong reject} \\
\addlinespace
\texttt{Respond with JSON only:} \\
\texttt{\{"rating": <one of 0, 2, 4, 6, 8, 10>, "confidence": <integer 1-5>, "justification": "<at most 50 words>"\}} \\
\bottomrule
\end{tabular}
\caption{The \textcolor{darkyellow}{$P_\text{reviewer}$} prompt, quoted verbatim, shown under open review with the \textcolor{darkyellow}{abstract and the introduction} as the content. It adds a reviewer persona and the six aspects on which the submission is to be judged, and it leaves the rating scale unchanged. The Author line is omitted under blind review and the prompt is otherwise byte-identical.}
\label{tab:llm_prompt_p2}
\end{table}

\begin{figure*}[!t]
\centering
\begin{minipage}{\textwidth}
\rule{\textwidth}{0.8pt}
\small
\begin{verbatim}
You are reviewing a submission to ICLR 2026. ICLR is a top-tier machine
learning conference, and it is highly selective, accepting roughly one in
three submissions. A paper accepted at ICLR is expected to present a novel
idea that advances the field, to support its claims with thorough
experiments that include strong baselines and ablations, and to be written
clearly. You are a senior expert with a background in artificial
intelligence, machine learning, natural language processing, and computer
vision. Judge the submission on its presentation, the extent of its
experimental evaluation, the quality of its writing, its clarity, its
soundness, and the strength of its core idea.

Title: {paper_title}
Author: {author_reputation_descriptor}.

Abstract: {paper_abstract}

Introduction: {paper_introduction}

Assign an overall rating on the ICLR 2026 scale. Ratings of 0, 2, and 4 are
negative and mean the paper should be rejected; ratings of 6, 8, and 10 are
positive and mean the paper should be accepted:

10 = strong accept; outstanding on every dimension, a clear highlight of
     the conference

8 = accept; strong experiments, clear writing, and a genuinely novel idea

6 = weak accept; solid experiments, good writing, and a reasonably novel
    idea

4 = borderline reject; no clear strength stands out, but nothing is clearly
    wrong either

2 = reject; major problems, for example in the experimental work

0 = strong reject; the submission reads like an unfinished draft

Respond with JSON only:
{"rating": <one of 0, 2, 4, 6, 8, 10>, "confidence": <integer 1-5>,
"justification": "<at most 50 words>"}
\end{verbatim}
\rule{\textwidth}{0.8pt}
\end{minipage}
\caption{The \textcolor{darkyellow}{$P_\text{both}$} prompt with the \textcolor{darkyellow}{abstract and the introduction} as the content, shown under open review. The Author line is omitted under blind review and the prompt is otherwise byte-identical.}
\label{fig:prompt-p3-text}
\end{figure*}

\begin{figure*}[htbp]
\centering
\begin{minipage}{\textwidth}
\rule{\textwidth}{0.8pt}
\small
\begin{verbatim}
You are reviewing a submission to ICLR 2026. ICLR is a top-tier machine
learning conference, and it is highly selective, accepting roughly one in
three submissions. A paper accepted at ICLR is expected to present a novel
idea that advances the field, to support its claims with thorough
experiments that include strong baselines and ablations, and to be written
clearly. You are a senior expert with a background in artificial
intelligence, machine learning, natural language processing, and computer
vision. Judge the submission on its presentation, the extent of its
experimental evaluation, the quality of its writing, its clarity, its
soundness, and the strength of its core idea.

Title: {paper_title}
Author: {author_reputation_descriptor}.

The main text of the submission is provided below as page images
(pages 1-9).

[page image 1] [page image 2] ... [page image 9]

Assign an overall rating on the ICLR 2026 scale. Ratings of 0, 2, and 4 are
negative and mean the paper should be rejected; ratings of 6, 8, and 10 are
positive and mean the paper should be accepted:

10 = strong accept; outstanding on every dimension, a clear highlight of
     the conference

8 = accept; strong experiments, clear writing, and a genuinely novel idea

6 = weak accept; solid experiments, good writing, and a reasonably novel
    idea

4 = borderline reject; no clear strength stands out, but nothing is clearly
    wrong either

2 = reject; major problems, for example in the experimental work

0 = strong reject; the submission reads like an unfinished draft

Respond with JSON only:
{"rating": <one of 0, 2, 4, 6, 8, 10>, "confidence": <integer 1-5>,
"justification": "<at most 50 words>"}
\end{verbatim}
\rule{\textwidth}{0.8pt}
\end{minipage}
\caption{The \textcolor{darkyellow}{$P_\text{both}$} prompt with the first \textcolor{darkyellow}{nine pages of the main text} supplied as rendered page images. Only the content block differs from Figure~\ref{fig:prompt-p3-text}. Each page is cropped to the type block so that the running header, the page number, and the line numbers are removed, and the author block on the first page is replaced by the anonymous placeholder of the ICLR template.}
\label{fig:prompt-p3-image}
\end{figure*}

\clearpage

\begin{figure*}[!ht]
\centering
\includegraphics[width=\linewidth]{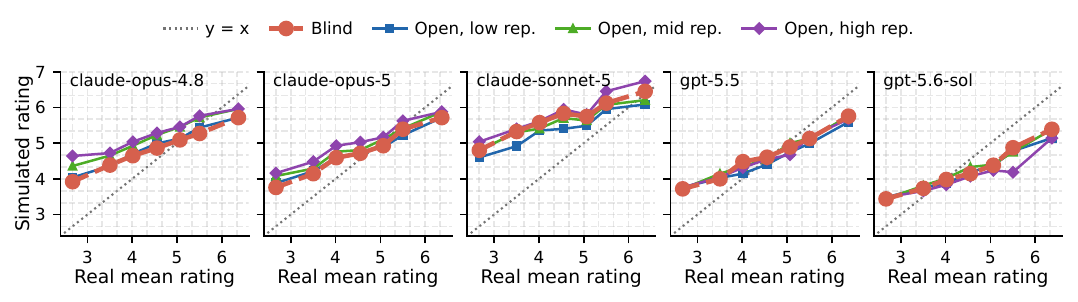}
\caption{Mean simulated rating against the real mean rating of the submission under the \textcolor{darkyellow}{$P_\text{conference}$} prompt, by author reputation condition, for \textcolor{darkyellow}{5  models}. The content is the \textcolor{darkyellow}{abstract and the introduction.} Submissions are grouped into the seven strata of the real rating. Both axes carry the ICLR rating scale and the dotted diagonal marks $y=x$.}
\label{fig:llm-sim-p1}
\end{figure*}

\begin{figure*}[!ht]
\centering
\includegraphics[width=\linewidth]{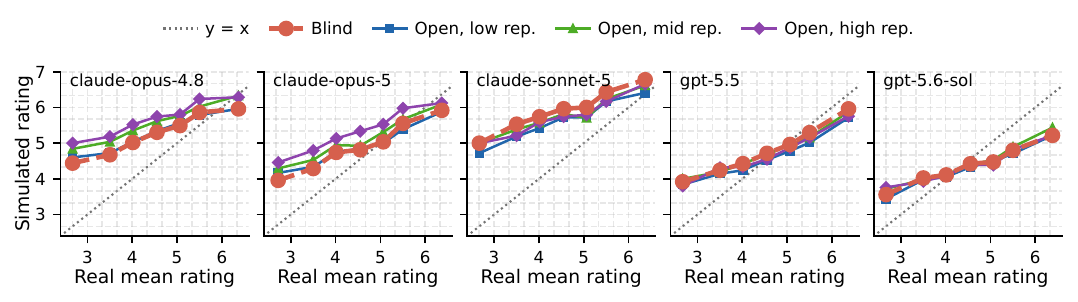}
\caption{Mean simulated rating against the real mean rating of the submission under the \textcolor{darkyellow}{$P_\text{reviewer}$} prompt, by author reputation condition, for \textcolor{darkyellow}{5  models}. The content is the \textcolor{darkyellow}{abstract and the introduction.} Submissions are grouped into the seven strata of the real rating. Both axes carry the ICLR rating scale and the dotted diagonal marks $y=x$.}
\label{fig:llm-sim-p2}
\end{figure*}

\end{document}